\documentclass[twocolumn,preprintnumbers,amsmath,amssymb,prl,floatfix]{revtex4-1}

\usepackage{dcolumn}
\usepackage{bm}
\usepackage{graphicx}
\usepackage{color}

\begin{document}

\title{Rare events in population genetics: Stochastic tunneling in a two--locus model
  with recombination}

\author{Alexander Altland, Andrej Fischer, Joachim Krug, and Ivan G. Szendro}

\affiliation{
Institut f\"ur Theoretische Physik, Universit\"at zu K\"oln,
D-50973 K\"oln, Germany}

\date{\today}

\begin{abstract}
We study the evolution of a population in a two-locus
genotype space, in which the negative effects of two single
mutations are overcompensated in a high fitness double mutant. 
We discuss how the interplay of finite population size, $N$,
and sexual recombination at rate $r$ affects the escape times $t_\mathrm{esc}$ to the double
mutant. For small populations demographic noise generates massive fluctuations in
$t_\mathrm{esc}$. The mean escape time varies non-monotonically with $r$,
and grows exponentially as $\ln t_{\mathrm{esc}} \sim N(r - r^\ast)^{3/2}$ beyond a critical value
$r^\ast$. 
\end{abstract}

\maketitle

Point mutations at different loci of the genome affect the fitness of
living organisms via complex intra-genomic correlations. These
interactions, known as \textit{epistasis} in 
population genetics, range from elementary pair-interactions to
extended patterns reflecting the structure of genetic networks \cite{Phillips2008}. 
An important form of epistasis termed \textit{reciprocal sign epistasis} occurs when
the deleterious effects of a single point mutation get
(over)compensated by the beneficial effects of a secondary mutation
\cite{Poelwijk2007,Poelwijk2011}. The abundance of such elementary motifs 
is expected from the nature of complementary base pair binding 
in RNA \cite{Stephan1996,Higgs1998}, but it also stands to reason that they
act as building blocks of evolutionary processes
in more complex 'fitness landscapes' \cite{Poelwijk2011,Weinreich2005a}.

The effects of epistatic pair correlation can be conveniently studied
in a two-locus
two-allele prototype system, i.e. a projection of the full genomic structure
onto just two loci with alleles $a,A$ and $b,B$,
respectively. 
Here, the term 'locus' refers to a specific genomic
position, and the 'alleles' $a,\dots,B$ stand for
the nucleotides present at the loci.
Throughout this paper we assume a fitness assignment such that 
the double mutant $AB$ is superior to
the wildtype $ab$ and both have higher fitness than the single mutants 
$Ab, aB$, whose fitnesses are moreover taken to be equal. 
When evolving under the joint influence of mutation, selection,
sexual recombination and demographic noise, this system exhibits dynamical
phenomena on a variety of time scales. 
Of particular interest here is the 'switching' process 
occurring at very large time scales, when the effects of a deleterious single mutation at one of the
two loci $ab\to Ab$ or $ab\to aB$ need to be overcome to reach a
high-fitness doubly mutated configuration $(Ab/aB)\to AB$.

Several decades of research on the two-locus system notwithstanding 
\cite{Stephan1996,Higgs1998,Crow1965,Eshel1970,Iwasa2004,Weinreich2005b,Weissman2009,Gokhale2009,Jain2010,Park2010,Lynch2010,Weissman2010},
important aspects of the above compensatory mutation mechanism remain
insufficiently understood. While the switching process, termed
\textit{stochastic tunneling} in the literature \cite{Iwasa2004}, is
relatively well understood in asexuals
\cite{Weissman2009,Gokhale2009}, the behavior induced by sexual
recombination is surprisingly complex. This is because of the dual
role played by recombination in this system. On the one hand,
recombination of the unfavorable single mutants $Ab+aB\to AB$ provides a channel of $AB$-generation that
does not rely on mutation. On the other hand, this mechanism competes with 
back-recombination $AB+ab\to Ab,aB$ which breaks up $AB$-genotypes
once they have formed. 

Detailed studies of the deterministic, infinite
population  dynamics show that the latter mechanism
categorically wins in large populations. 
Beyond a critical value $r^\ast$ of the recombination rate $r$
a stable stationary solution localized at the wildtype genotype $ab$ 
emerges \cite{Park2010},
and the escape to the double mutant $AB$ is completely suppressed
\cite{Jain2010}. However, computer simulations of finite populations
also display a parameter regime for $r < r^\ast$ where the escape is aided by
recombination, such that 
an initial decrease of the escape time $t_\mathrm{esc}$ eventually
gives way to an increase at larger values of $r$ \cite{Weinreich2005b}.  
The effect of recombination on the speed of adaptation is thus seen to
depend on the population parameters in a complex way, as has also been
observed in studies using empirical fitness data \cite{deVisser2009}.

In addition to the recombination rate, the dynamics of the two-locus
system is governed by the population size $N$, the mutation rate $\mu$,
and the strength of selection given by the typical scale of
fitness differences $s$. Here we focus on populations that are 
moderately large, in the sense that $N \mu, Ns \gg 1$, and subject to
strong selection with $s/\mu \gg 1$ \cite{footnote1}.   
Using a linearization of the full master equation near the
initial wild type population, we derive an approximate expression for the full
distribution of escape times for $r < r^\ast$. This allows us to identify two
fundamentally different switching scenarios. In populations that are  
smaller than a characteristic size to be specified below,  
the temporal bottleneck of the evolution is the appearance of 
the first few individuals of the $AB$ populations, which is a rare
event with an exponentially distributed waiting time. The most likely
escape time is then much smaller than the
typical time, and fluctuations in $t_\mathrm{esc}$ are of the same
order as the mean. In contrast, in large populations the evolution is
limited by the growth of the $AB$ population and the distribution of
$t_\mathrm{esc}$ is sharply peaked around the mean. The
  nonmonotonicity of $t_\mathrm{esc}$ observed in
  \cite{Weinreich2005b} is a feature of the fluctuation-dominated
  regime which disappears in large populations. 

For $r > r^\ast$ the emergence of bistability in the deterministic
dynamics implies that the problem becomes similar to the noise-driven 
escape from a metastable state. For this kind of problem path-integral methods
akin to semi-classical quantum mechanics have recently been developed
\cite{Assaf2010}, and we show that they can be applied in the present
context as well.

\textit{Model.} We consider the dynamics of the two-locus system, as
governed by the interplay of selection, mutation and 
recombination. The two low fitness genotypes $Ab, aB$ are lumped into a single
subpopulation. This is justified when the creation rate of single
mutants satisfies $N \mu \gg 1$, such that the number of single mutant
individuals is large compared to unity at all times
and the recombination process $Ab + aB
\longrightarrow (ab/AB)$ is not limited by the simultaneous presence of
both parental types.  
The fitness of the three types is given by,
respectively, $ab\leftrightarrow 1$, $(Ab/aB)\leftrightarrow 1-s_d$,
$AB\leftrightarrow 1+s_b$, where $s_{d,b}$ denote the selection
coefficients of the deleterious single mutants and the beneficial
double mutant relative to the wild type $ab$, respectively.
Mutation alters allelic content at a rate $\mu \ll s_d,s_b$. For
instance, $ab \stackrel{2\mu}{\longrightarrow} Ab/aB$, where the
factor of $2$ accounts for the fact that the change of either allele,
$a$ or $b$ generates a single mutant. For simplicity, we will neglect
the effect of back mutation $Ab/aB\to ab$ in our analysis of the
escape time. In view of the smallness of the single mutant population
(see below), this assumption is largely inconsequential. Finally, the
random mating of individuals generates offspring by recombination at a
rate $r$; for example, $AB +
ab\stackrel{r}\longrightarrow (Ab/aB)$.

The (Moran \cite{Durrett2002}) master equation for the evolution of the three
population sizes, $n_0\leftrightarrow ab, n_1 \leftrightarrow Ab+aB$,
and $n_2\leftrightarrow AB$ at constant total population
$N=n_0+n_1+n_2$ reads as $\partial_t P({\bf x},t) = H({\bf x}) P({\bf x},t)$, where
${\bf x} = (x_0, x_1, x_2)$, 
\begin{widetext}
  \begin{align}
\label{eq:1}
     H({\bf x})=\sum_{i=0}^2 \left[(E_i^+-1) d_i({\bf x}) +
      (E_i^--1)b_i({\bf x})+(E_i^+E_{i+1}^- -1)m_i({\bf x})\right]+\sum_{i,j=0}^2
    (E_i^+E_j^- -1) r_{ij}({\bf x}),
  \end{align}
\end{widetext}
and the notation emphasizes the analogies to an imaginary time
'Schr\"odinger equation'. In (\ref{eq:1}), $x_i=n_i/N$ are the
genotype 'frequencies', the 'operators' $E_i^\pm f(x_i)\equiv e^{\pm
  {1\over N} \partial_{x_i}}f(x_i) =f(x_i \pm N^{-1})$ act by
translation by one individual, and the rates $d_i, b_i, m_i,r_{ij}$
are the death, birth, mutation, and recombination
rates~\cite{rate_fn}.  Equation~(\ref{eq:1}) describes the
evolution of the system at one individual death/birth event per time
step. Alternatively, one may employ a Wright-Fisher approach where an
entire generation update is performed at each step
\cite{Durrett2002}. While this is numerically faster by a factor of
${\cal O}(N)$, and has been used in parts of our computer simulations,
the long rangedness of the Wright-Fisher equation in the variables
$(n_0,n_1,n_2)$ makes it difficult to handle analytically.

\begin{figure}[th]
  \centering
  \includegraphics[width=7cm]{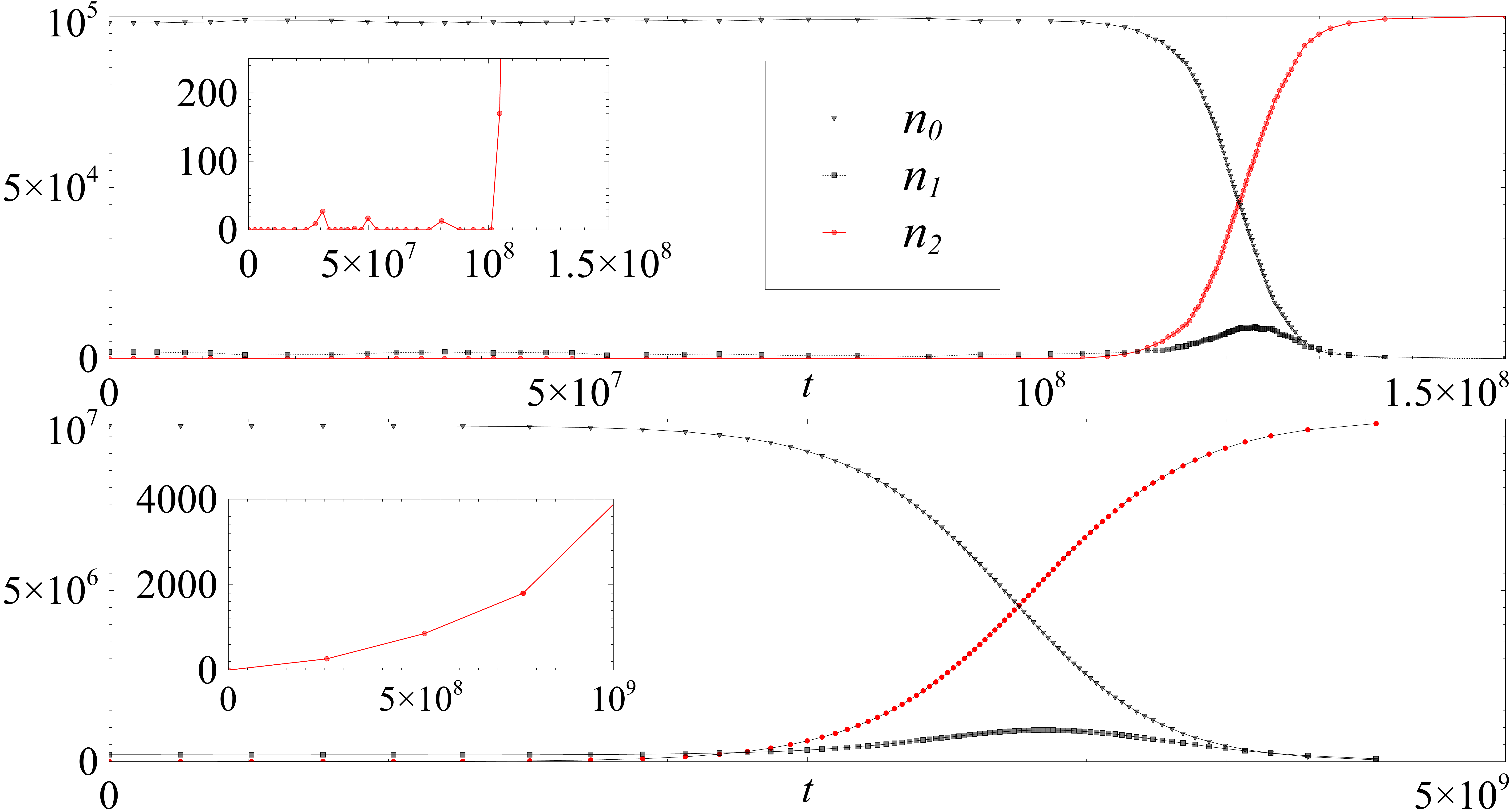}
  \caption{(Color online) Single runs of the escape process. Top ($\gamma N=0.2$): creation of the first
    $AB$-mutant determines escape time,
    bottom ($\gamma N=20$): sweep through bulk of the population
    determines escape time. In the insets the population size axes have been enlarged to 
highlight the initial stages of the process.}
   \label{fig:bottleneck} 
\end{figure}

\textit{Escape time.} Figure~\ref{fig:bottleneck} shows results for
the time dependence of $(n_0,n_1,n_2)$ obtained by simulation of the
master equation (\ref{eq:1}). These profiles depend sensitively on the
value of the parameter $ N \gamma$, where the meaning of the rate
$\gamma=2{\mu^2\over s_d} + r \left({\mu\over 
    s_d}\right)^2$ will be discussed momentarily. 
For $\gamma N\ll 1$ (upper panel), the limiting factor
for the escape process is the stochastic generation of a sufficient
number of $AB$-individuals (cf. the inset). Once the $AB$-clone
reaches a critical size, the fitness advantage
triggers a fast sweep through the entire population. In the opposite
case $\gamma N \gg 1$, the  quasi-deterministic increase of the
$x_2$-population determines the escape time. 

We aim to describe
the dynamical processes  characterizing the respective temporal bottlenecks
from the master equation. 
To this end, we first notice that starting from an initially
un-mutated 'wild population' $(x_0,x_1,x_2)=(1,0,0)$, a competition of
mutation out of the $ab$-population and a counter-acting selection
pressure $s_d$ generates a population $x_1\equiv \bar x_1 =
2\mu/s_d$ at short time scales $\sim {s_d}^{-1}$. In the parameter regime
of interest here
there are many individuals in a singly mutated state at any instance
of time, $\bar x_1 \gg N^{-1}$.
During the evolutionary stages determining the switching time, the
frequencies $x_0\simeq 1$ and $x_1\simeq \bar x_1$ do not change
significantly and the assumption of independence of $P$ on $(x_0,x_1)$
does not lead to qualitative errors. We thus substitute
$(x_0,x_1)\simeq (1,\bar x_1)$ into (\ref{eq:1}) and set $x_2\equiv
x\equiv n/N$
for notational simplicity to obtain the linear master equation $\partial_t P(n,t) = H(n) P(n,t)$, where
\begin{align}
  \label{eq:2}
  H(n) = (E^--1) (R_+ n+\gamma)+(E^+-1)R_- n
\end{align}
and $R_+={1\over N}(1+s_b+r/2)$, $R_-={1\over N}(1+3r/2)$.
The coefficient $\gamma$ has been defined above, and can now 
be understood as the effective rate of generation of double mutants, including 
contributions from both the mutational  ($\sim \mu \, \bar{x}_1$) and 
recombinational channel ($\sim r \, \bar{x}_1^2$). The usage of the
discrete variable $n\in \Bbb{N}$ in (\ref{eq:2}) accounts for the importance
of the 'quantization' of individuals to our further analysis. Starting
from zero $AB$-individuals, $P(n,t=0)=\delta_{n,0}$, the time scale for
the generation of an $AB$-clone is set by $\gamma^{-1}$. 

To obtain $P(n,t)$ we adapt the general solution \cite{vanKampen} for the generating
function $G(z,t) =\sum_{n=0}^\infty z^n P(n,t)$ of linear master equations to
Eq.~(\ref{eq:2}). This yields $G(z,t) = (1+\kappa(t)(1-z))^{-\lambda}$, where
$\lambda=\gamma/R_+ \sim N \gamma$, $\kappa(t) = {R_+\over
  \Delta R}(e^{\Delta R t}-1)$, and
$\Delta R=R_+-R_- > 0$. Computing the inverse transform,
we obtain 
\begin{align}
\label{eq:3}
P(n,t) =
{\kappa^n\over (\kappa + 1)^{n+\lambda} }\prod_{j=0}^{n-1}{j+\lambda\over
  j+1}.
\end{align} 
For fixed $n$, $P(n,t \to \infty)$ scales to zero showing
that the distribution flows towards large values of $n$.
Writing the master equation in the form $\partial_t P = J_{n-1} -
J_n$, we identify  
the current $J_n(t) = (R_+ n + \gamma) P(n,t) - R_- (n+1)
P(n+1,t)\simeq \Delta R n P(n,t)$ at which probability flows 
through a fixed reference value $n \gg 1$.
The flow of the distribution
implies $\int dt \, J_n(t) =1$, which shows that $J_n(t)$ may be
interpreted as the distribution $f_n(t)$ of escape times through $n$.  
We here consider $n\simeq
N/2$ as a reference value where the $AB$-population is about to take
over, while the linearized approximation of the master equation has
not yet become fully invalid. (For most parameter values, the ensuing
threshold times are comparable to the times of full fixation, $n=N$.)
For large $n$ the product in (\ref{eq:3}) can be expressed in terms of the
$\Gamma$-function, which yields the result
\begin{align}
\label{eq:4}
  f_n(t) \approx \Delta R {\kappa(t)^{n}\over (\kappa(t)+1)^{n+\lambda}}
  n^\lambda \Gamma^{-1}(\lambda).
\end{align}
Two time scales can be extracted from (\ref{eq:4}). For $t \to \infty$
the distribution decays exponentially on the time scale 
$t_\mathrm{tail} = {R_+\over  \gamma \Delta R}$. The second time scale
is the time at which (\ref{eq:4}) is maximal, given by
$t_\mathrm{max} = (\Delta R)^{-1} \ln(\Delta R n/\gamma+1)$. This is
also the time when the solution of the rate equation 
$\dot{n} = \Delta R n + \gamma$ corresponding to (\ref{eq:2})
reaches the value $n$. Apart from the logarithmic factor we see
that $t_\mathrm{max}/t_\mathrm{tail} = \lambda$, showing that the parameter $\lambda$
distinguishes the two dynamic regimes described above. For $\lambda
\ll 1$ the distribution (\ref{eq:4}) becomes purely exponential,
and fluctuations in
$t_\mathrm{esc}$ are of the same order as the mean $t_\mathrm{tail}$. In contrast,
for $\lambda \gg 1$ the distribution is sharply peaked around the most
likely value $t_\mathrm{max}$, with fluctuations of order $\Delta t$ with 
$\Delta t/t_\mathrm{max} \sim N^{-1/2}$. 

\begin{figure}[b]
  \centering
  \includegraphics[width=7cm]{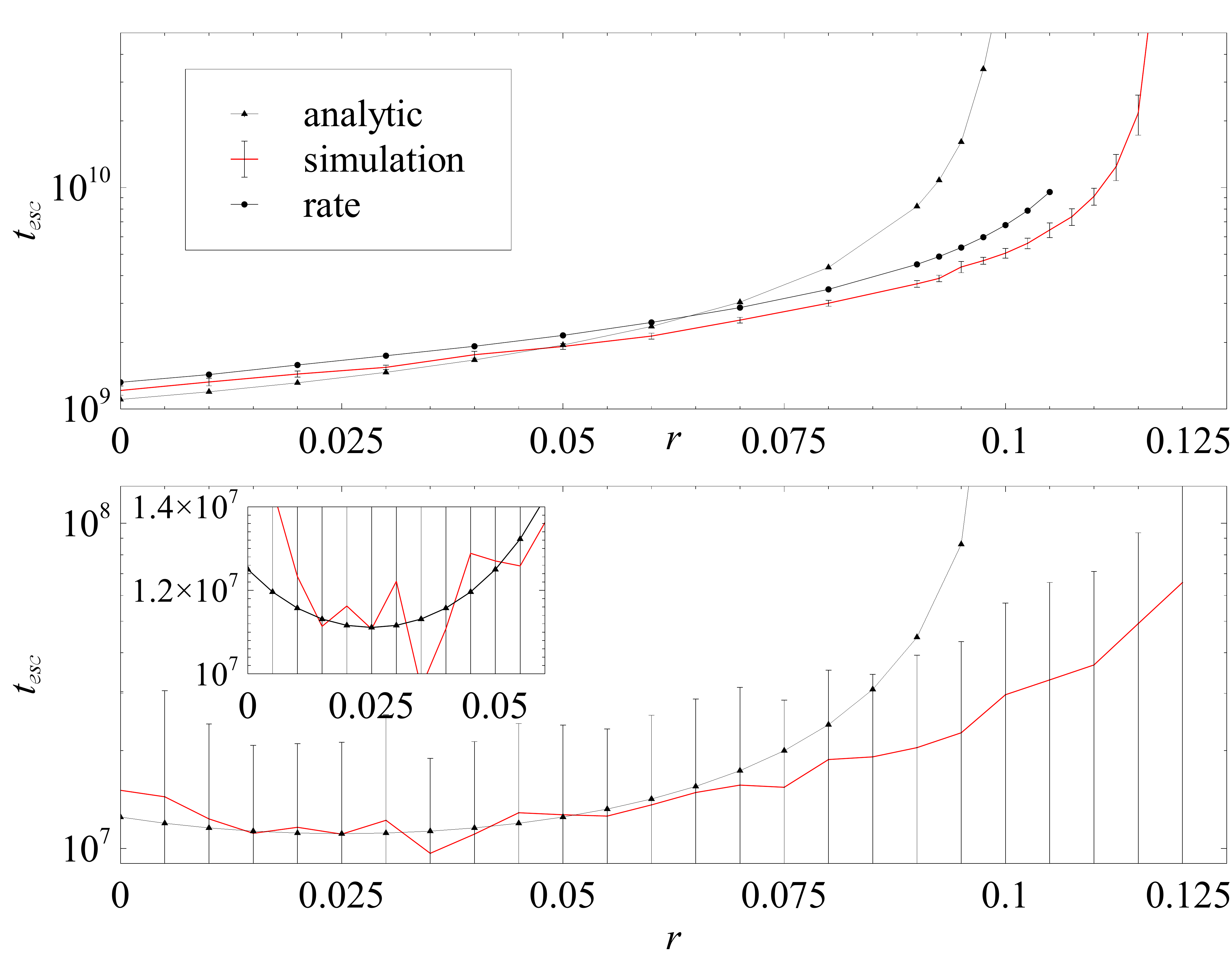}
  \caption{(Color online) Mean escape times as a function of $r$
    for  $s_b=0.1, s_d=0.025, \mu=10^{-4}$. Upper panel: $N=10^7$,
    lower panel, $N=10^4$. Error bars in lower panel illustrate the
    large fluctuations in the escape time, and inset highlights the minimum
  in $t_\mathrm{esc}$ at $r = r_\mathrm{min}$.}
   \label{fig:t_esc} 
\end{figure}

Inserting the explicit expressions for $R_+$, $\Delta R$ and $\gamma$
it is straightforward to show that $t_\mathrm{tail}$ displays a minimum as a
function of $r$ provided $s_d < s_b/2$, i.e. in cases where the
fitness valley separating the genotypes $ab$ and $AB$ is
relatively shallow. The minimum is located at $r_\mathrm{min} = s_b/2-
s_d < s_b/2$. The maximal speedup due to recombination can be substantial, and is given by
$t_\mathrm{tail}(r_\mathrm{min})/t_\mathrm{tail}(r=0) \approx \frac{8 s_d}{s_b}$ for $\mu \ll s_d \ll
s_b$ \cite{footnote2}. By contrast, $t_\mathrm{max}$ increases monotonically with $r$.

\textit{Metastability.}
Both time scales diverge when $\Delta R \sim (s_b - r)$ vanishes as $r \to s_b$. Within the simplified model
defined by (\ref{eq:2}), this reflects the emergence of a stable
stationary distribution centered around the wildtype genotype $ab$,
which has been found in previous studies of the deterministic model 
\cite{Eshel1970,Park2010,Jain2010}. 
Figure \ref{fig:t_esc} shows the mean values $t_\mathrm{esc}(r)$
(triangles) along with the escape times obtained from the simulation
of the Moran master equation (\ref{eq:1}) (bars), and from the solution of
the corresponding
deterministic rate equations (dots). All three curves show a rapid
increase of the escape time at the critical value $r^\ast\simeq s_b$.
The rationale behind this behavior is that at strong recombination 
the  'reshuffling' $AB+ab\stackrel{r}{\to}
Ab,bA$, outperforms the fitness advantage of the $AB$ population,
thus preventing its growth.

We sketch the analysis of the model near criticality, $r\simeq
r^\ast$. To this end we consider a Fokker-Planck
approximation to (\ref{eq:1}), i.e. a second order expansion in the
'momentum' variables $p_i=\partial_{x_i}$. This results in an operator of the generic form 
$H=-{1\over N} p_i F_i + {1\over 2N^2} p_{i} X_{ij} p_j$, where the $F_i$ denote the components of the 
deterministic drift term and $X_{ij}$ is the diffusion matrix. Eliminating one frequency by normalization,
$x_0=1-x_1-x_2$, the problem becomes two-dimensional with $i=1,2$. 
The quickly equilibrating frequency $\bar x_1 \sim \frac{2\mu}{s_d}$ of the
single mutant population follows from
stationarity under drift, $F_1(x_0,\bar x_1)=0$, which gets
us to the effective one-dimensional Hamiltonian, $H=-{1\over N}p F +
{1\over 2N^2} Xp^2$, where we have set $x_2=x$ and $p_2=p$ for
notational simplicity. Then $F(x)\equiv F_2(x,\bar x_1)$, $X(x)\equiv
X_{22}(x,\bar x_1) \approx x\,(1-x)$ describe the
effective dynamics of the $AB$-population.

\begin{figure}
  \centering
  \includegraphics[width=7cm]{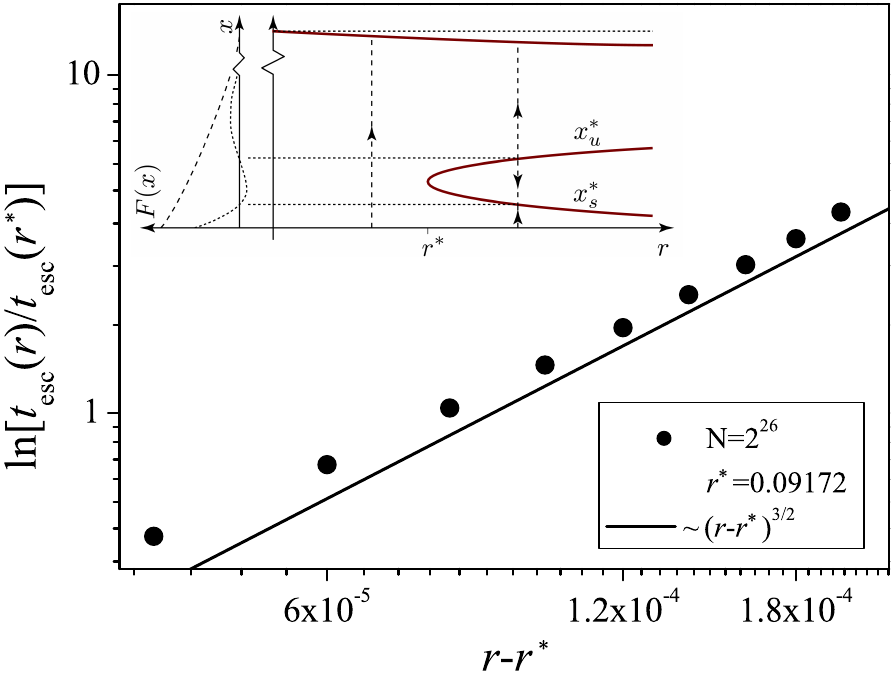}
  \caption{(Color online) Main figure shows the logarithm of the escape time,
    normalized to its value at $r = r^\ast$, as a function of
    $r-r^\ast$ in double-logarithmic scales. The escape time was
    defined as the time when the $AB$-clone reaches
    80\% of the total population. Black dots correspond to
simulations of a population of size $N = 2^{26} \approx 6.7 \times
10^7$ with parameters $s_b= 0.1$, $s_d =0.025$, and $\mu=10^{-5}$.  
All data are averaged over $10^5$ runs.
The straight line has slope $3/2$. Inset illustrates the
deterministic one-dimensional dynamics of the $AB$-population. The vertical graph on the left 
shows the shape of the function $F(x)$ for $r < r^\ast$
(dashed line) and $r > r^\ast$ (dotted line). The horizontal graph shows the evolution
of the fixed point structure with $r$. For $r < r^\ast$ the only fixed point
is near $x_2 = x \approx 1$; for $r > r^\ast$ another pair of
fixed points ($x_s^\ast, x_u^\ast$) emerges near $x \approx 0$.}
   \label{fig3} 
\end{figure}

The singularities in the escape times reflect
the emergence of a pair of stable ($x^\ast_s$) and unstable
($x^\ast_u$) fixed points, $F(x^\ast_{s,u})=0$, $0<x^\ast_s<x^\ast_u\ll 1$, close
to the boundary $x=0$ (inset of Fig.\ref{fig3}). The unstable fixed point $x^\ast_u$ defines a
'recombination barrier' \cite{Stephan1996} which  
blocks the fixation of the fitter population out of a small number of 
initial individuals $x(0) <x^\ast_u$. Based on general results for the
escape from metastable states obtained within a large deviations ('WKB')
approach \cite{Assaf2010}, one expects the escape
time to grow exponentially with $N$ as $t_\mathrm{esc} \sim \exp[C N
\delta^3]$, where $C > 0$ is a constant and $\delta \sim x^\ast_u -
x^\ast_s$. The analysis of the deterministic
Wright-Fisher version of the problem shows that $\delta \sim (r -
r^\ast)^{1/2}$ \cite{Park2010}, and we predict that
$t_\mathrm{esc} \sim \exp[C N (r - r^\ast)^{3/2}]$. 
Figure~\ref{fig3} shows data for the escape time obtained 
by simulating a Wright-Fisher process~\cite{deVisser2009} at 
values of $r$ slightly above $r^*$. The value $r^*\simeq 0.09172$ was
calculated using expressions derived in \cite{Park2010}
for the deterministic ($N \to\infty$) limit. The results are seen to be in good
agreement with the prediction $\ln t_\mathrm{esc} \sim (r -
r^\ast)^{3/2}$. 

\textit{Conclusion.}
To summarize, we have presented an analysis of a
paradigmatic two-locus model of population genetics. We have seen how,
depending on the population size, 
recombination may speed up or
delay the evolution towards the high fitness state. A
challenge for future work is to develop the WKB-type approximation for
$r > r^\ast$ into a fully quantitative theory, which can be used to
predict the size of the recombination barrier in a specific biological
setting.

This work was supported by DFG within SFB TR 12 and SFB 680.
We thank U. Gerland, B. Meerson, D.B. Weissman and R. Neher for useful 
discussions and remarks.


\begin{thebibliography}{20}
\bibitem{rate_fn} The rates in (\ref{eq:1}) are defined as follows: Death
  rates $d_i = x_i$; birth
  rates $b_i = (w_i/\bar{w})x_i$, where $w_i$ are the genotype
  fitnesses given in the text and $\bar{w} = \sum_{i=0}^2 w_i x_i$ is
  the population mean fitness; mutation rates $m_0 = 2 \mu x_0$, $m_1
  = \mu x_1$; and recombination rates $r_{01} = r x_0 (x_1+x_2)$, 
$r_{02} = r_{20} = (r/2)x_0 x_2$, $r_{10} = r x_0 x_1 + (r/4) x_1^2$, $r_{12} =
r x_1 x_2 + (r/4) x_1^2$, $r_{21} = r x_2(x_0 + x_1 )$.
\end{thebibliography}
\end{document}